\documentclass[aip,jap,preprint,groupeauthor,showpacs]{revtex4-1}

\usepackage{amsmath}
\usepackage{graphicx}
\usepackage{dcolumn}
\usepackage{bm}
\usepackage[dvips]{color}
\usepackage{ulem}

\begin{document}

\title{Optical properties and electronic structure of multiferroic hexagonal orthoferrites $R$FeO$_3$ ($R$=Ho, Er,Lu)}

\author{V.\,V. Pavlov} \affiliation{Ioffe Physical-Technical Institute, Russian Academy of Sciences,
194021 St. Petersburg, Russia} 
\author{A.\,R. Akbashev} \affiliation{Materials Science Department, Moscow State University, 119991 Moscow, Russia}
\author{A.\,M. Kalashnikova}\affiliation{Ioffe Physical-Technical Institute, Russian Academy of Sciences,
194021 St. Petersburg, Russia}
\author{V.\,A. Rusakov}\affiliation{Ioffe Physical-Technical Institute, Russian Academy of Sciences,
194021 St. Petersburg, Russia}
\author{A.\,R. Kaul}\affiliation{Chemistry Department, Moscow State University, 119991 Moscow,
Russia}
\author{M. Bayer}\affiliation{Ioffe Physical-Technical Institute, Russian Academy of Sciences,
194021 St. Petersburg, Russia}\affiliation{Experimentelle Physik 2, Technische Universit\"at
Dortmund, 44221 Dortmund, Germany}
\author{R.\,V. Pisarev}\affiliation{Ioffe Physical-Technical Institute, Russian Academy of Sciences,
194021 St. Petersburg, Russia}

\date{\today}

\begin{abstract}
We report on optical studies of the thin films of multiferroic hexagonal (P.G.\,6\textit{mm}) rare-earth
orthoferrites $R$FeO$_3$ ($R$=Ho, Er, Lu) grown epitaxially on a (111)-surface of
ZrO$_2$(Y$_2$O$_3$) substrate. The optical absorption study in the range of
0.6-5.6\,eV shows that the films are transparent below 1.9\,eV;
above this energy four broad intense absorption bands are
distinguished. The absorption spectra are analyzed taking into account the unusual fivefold coordination of the Fe$^{3+}$ ion. Temperature dependence of the optical absorption at 4.9\,eV shows anomaly at 124\,K, which we
attribute to magnetic ordering of iron sublattices.
\end{abstract}

\maketitle

In the last decades materials based on transition-metal (TM) and rare-earth-metal (RE)
oxides have attracted much attention because of their importance for
multiple applications as well as for fundamental studies.\cite{Rao} A rich variety of interesting physical properties of TM oxides is determined by different types of TM ion and their local coordinations. The distinct place among the TM oxides belongs to the compounds based on iron ions, as they possess the largest magnetic moments and highest temperatures of magnetic ordering. It is well known that iron ions in oxides prefer to occupy either octahedral or tetrahedral positions and the examples of such materials are numerous.\cite{Rao,Cornell} By contrast, the five-fold coordination of iron ions is unstable and only recently several groups reported successful synthesis of powders, fine particles and thin films of hexagonal rare-earth orthoferrites $R$FeO$_3$.\cite{Yamaguchi,Mizoguchi,Bossak3,Nagashio,Magome-JJAP2010,Iida}

This new type of ferrites remains scarcely investigated.\cite{Akbashev-Review} Such an important property as multiferroicity of these compounds was under question till very recently. In early studies no magnetic ordering was identified in ultrafine hexagonal EuFeO$_3$ and YbFeO$_3$ particles.\cite{Mizoguchi} On the contrary, ferrimagnetic ordering for both iron and rare-earth sublattices was later proposed in hexagonal YbFeO$_3$ thin films,\cite{Iida} while recent magnetic measurements have indicated a weak ferromagnetic ordering in hexagonal $R$FeO$_3$ ($R$=Lu, Er-Tb)\cite{Akbashev-APL2011} and YbFeO$_3$ films\cite{Jeong-JACS2012} below the N\'{e}el temperature $T_N\sim$120\,K. Several claims of ferroelectricity in hexagonal $R$FeO$_3$ were made\cite{Bossak3,Iida,Magome-JJAP2010} and confirmed recently by the comprehensive studies of electric polarization in YbFeO$_3$, which revealed two ferroelectric transitions at 470 and 225\,K.\cite{Jeong-JACS2012} Despite of the progress in understanding the magnetic and ferroelectric properties of hexagonal $R$FeO$_3$, until now no theoretical or experimental investigations of their electronic structure has been reported.

In this Communication we report on the studies of optical properties and electronic structure of thin epitaxial
films of hexagonal $R$FeO$_3$ ($R$=Ho, Er, and Lu). In particular, optical absorption of these films was studied in the range of 0.6-5.6\,eV and analyzed in terms of the crystal field
theory.\cite{Burns,Lever} This allowed us to disclose the features of electronic structure of hexagonal orthoferrite related to the Fe$^{3+}$ ion in the unusual coordination. Furthermore, temperature study of absorption suggests an onset of magnetic ordering of iron sublattices at 124\,K, in agreement with recent magnetic studies.

Epitaxial films of hexagonal iron oxides
$R$FeO$_3$ ($R$=Ho, Er, Lu) were grown by metal-organic chemical vapor deposition (MOCVD) on single-crystalline
(111)-substrates of cubic yttria-stabilized zirconia
ZrO$_2$(Y$_2$O$_3$) (YSZ) using $R$(thd)$_3$  and Fe(thd)$_3$
as volatile precursors, where thd is
2,2,6,6-tetramethylheptane-3,5-dionate. The deposition temperature
was 900$^\mathrm{o}$C, the precursor evaporator temperature
250$^\mathrm{o}$C, the total gas pressure in the reactor was
7\,mbar, and the partial oxygen pressure 3.5\,mbar. The deposition
rate was 5\,nm/min and the final thickness of films was
50-70\,nm.

\begin{figure}[t]
\includegraphics[width=8.0cm,angle=0]{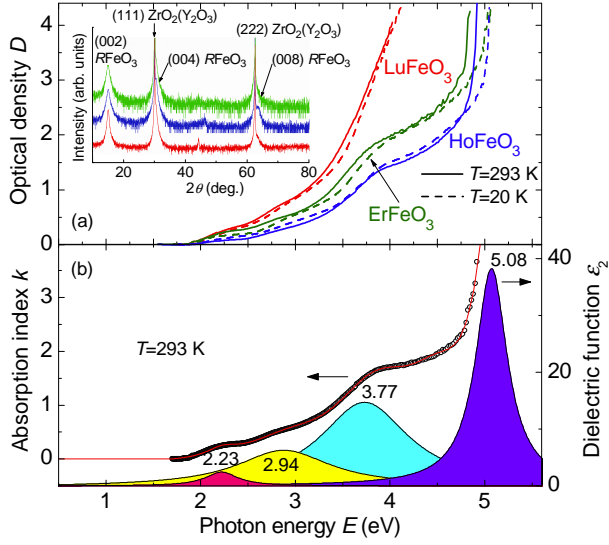}
\caption{(Color online) (a) Experimental optical density spectra of
hexagonal $R$FeO$_3$ ($R$=Ho, Er, Lu) films at $T=293$\,K
(solid lines) and $T=20$\,K (dashed lines). No corrections for
reflection losses were taken. Inset shows 2$\theta/\omega$ scans.
(b) Optical spectrum of the mean value of the absorption indexes of
HoFeO$_3$ and ErFeO$_3$ at $T=293$\,K (symbols) and its fit using
Eq.\,\ref{eq-Lorentz} (line). Color-shaded areas show single Lorentz
oscillators as obtained from the fit.} \label{fig2}
\end{figure}

A detailed X-ray diffraction (XRD) study using 4-circle
diffractometer revealed a highly epitaxial growth resulting in (0001)-oriented films. A characteristic (002)-reflection
of the hexagonal 6$mm$ phase was found in the 2$\theta/\omega$-scans
for all samples (see Inset in Fig.\,\ref{fig2}(a)). The in-plane XRD
study revealed (300)-reflections in 2$\theta$/$\chi$$\varphi$-scans
at (220)-YSZ reflection accompanied sometimes by a weak (110)-peak
indicating $<$110$>$ preferable in-plane orientation
with respect to the $<$110$>$ axis of YSZ. The corresponding 6 reflections were observed in $\varphi$-scans. Scanning electron
microscopy and atomic force microscopy studies showed a very smooth
surface of the films with the RSM (root mean square) roughness of 1-2\,nm over the
10$\times$10\,$\mu$m$^2$ area.

Fig.\,\ref{fig2}(a) shows optical density spectra of the films
at temperatures $T$=293 and 20\,K. All samples show negligible absorption below $\sim$1.9\,eV. Above
this energy the absorption begins to increase gradually and could
not be measured above $\sim$4.8\,eV at $T=293$\,K and above
$\sim$5.1\,eV at $T=20$\,K. The averaged absorption index $k$ for
HoFeO$_3$ and ErFeO$_3$ at room temperature is shown in
Fig.\,\ref{fig2}(b). The uncertainty in the films thickness yields
the accuracy for the absorption index of $\pm$30\%. The absorption spectrum of hexagonal $R$FeO$_3$ differs
significantly, on one hand, from those of perovskite orthoferrites
and related iron oxides,\cite{Kahn,Pisarev-PRB2009}
and, on the other hand, from hexagonal
manganites.\cite{Souchkov,Kalashnikova}

In order to identify the position of the absorption bands contributing to the spectra in Fig.\,\ref{fig2}
we decomposed the absorption index spectrum to the set of Lorentz oscillators according to
\begin{equation}
k=\mathrm{Im}\sqrt{\varepsilon};\,\,\varepsilon = a + bE + icE +\sum_{n}\frac{f_{n}}{E_{n}^{2}-E^{2}-iE\gamma_{n}}, \label{eq-Lorentz}
\end{equation}
where $\varepsilon$ in the complex dielectric function, $E$ is the photon energy; $a$, $b$ and $c$ are coefficients
describing collective contribution from all transitions above the
experimental spectral range. $E_{n}$, $f_{n}$ and $\gamma_{n}$ are
the resonance energy, oscillator strength and relaxation parameter
of the $n$-th oscillator, respectively. This analysis yielded four broad bands in the range above 1.9\,eV
as shown Fig.\,\ref{fig2}(b).

\begin{figure}[t]
\includegraphics[width=8.0cm,angle=0]{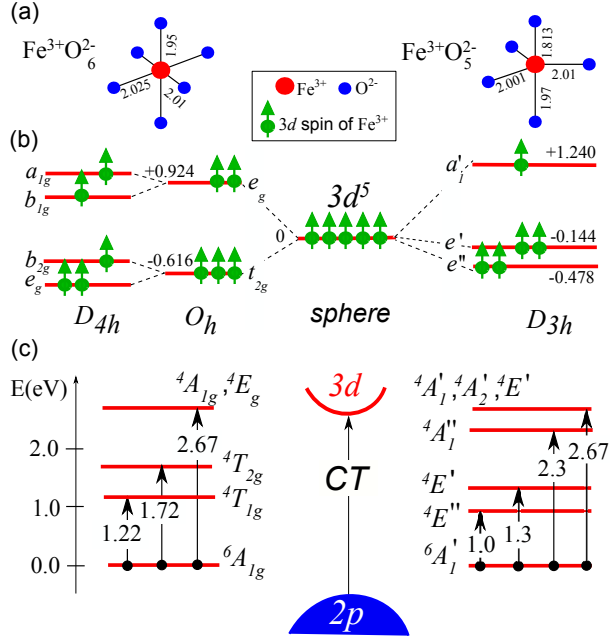}
\caption{(Color online) (a) Schematic presentation of the
Fe$^{3+}$O$^{2-}_6$ complex in a perovskite orthoferrite and the
Fe$^{3+}$O$^{2-}_5$ complex in a hexagonal orthoferrite. The bond
lengths are given in \AA.\cite{Zhou-PRB2008,Magome-JJAP2010} (b) Splitting and
occupation of the Fe$^{3+}$ ion ground state in the
spherically-symmetric, octahedral ($O_h$), distorted octahedral
($D_{4h}$) and trigonal-bipyramidal ($D_{3h}$) complexes. Numbers
show the energies of the orbitals (in eV) with respect to the ground state
in the spherical coordination. (c) Ground and the four lowest excited states of Fe$^{3+}$ in distorted octahedron and
trigonal bipyramid. Arrows with numbers show the CF transitions and their energies (in eV).
Also shown (in the middle) is the lowest CT transition from the
valence to the conduction band.}\label{Fig-SpectralLevels}
\end{figure}

We analyze the possible origin of the observed absorption
bands in hexagonal $R$FeO$_3$ on the basis of the crystal field
(CF) theory for the 3$d^5$ ion with octahedral and trigonal bypiramidal oxygen coordinations, shown in Fig.\,\ref{Fig-SpectralLevels}(a).\cite{Lever} The ground state of the Fe$^{3+}$ ion is the high-spin singlet (Fig.\,\ref{Fig-SpectralLevels}(b)), with any crystal field excitation being a transition between majority- and minority-spin 3$d$ orbitals. The energies of the excited states in the Fe$^{3+}$O$^{2-}_{5,6}$ complex (Fig.\,\ref{Fig-SpectralLevels}(c)) can an be
expressed in terms of the crystal field parameters $\Delta_0$, $Ds$ and $Dt$ and exchange
splitting parameter $\varepsilon_0$.\cite{Lever,Palacio-JChemPhys1976} We
calculated the CF parameters in the Fe$^{3+}$O$^{2-}_5$ complex in hexagonal $R$FeO$_3$ using the known values of these energies for the octahedral complex Fe$^{3+}$O$^{2-}_6$ in perovskite
orthoferrites (Fig.\,\ref{Fig-SpectralLevels}).\cite{Usachev-PhysSolSt2005} Here we took an advantage of the unequivocal relation between $\Delta_0$ parameters in
different Fe$^{3+}$O$^{2-}_N$ complexes\cite{Burns}
and the dependence of the CF parameters on the Fe$^{3+}$-O$^{2-}$
bond lengths.\cite{Lever,Pisarev-PRB2011} The resulting CF
parameters for the hexagonal orthferrite are $\Delta_0=1.76$\,eV, $Ds=$0.149\,eV and $Dt=0.157$\,eV.
The exchange splitting $5\varepsilon_0$ between the majority- and minority-spin orbitals defines solely the transition
$^6A'_{1}\rightarrow\,^4A'_1,\,^4A'_2,\,^4E'$ and is similar in a vast
majority of iron oxides (see e.g. Ref.\,\onlinecite{Burns},
Table\,5.15). Therefore, we take this energy to be close to that in the perovskite orthoferrite (2.67\,eV).\cite{Usachev-PhysSolSt2005} Knowing this value and the crystal field parameters $Ds$
and $Dt$ we calculated the energies for other three low lying
excited states, as shown in Fig.\,\ref{Fig-SpectralLevels}(c).

No features in the absorption spectra of hexagonal orthoferrites are observed below 1.9\,eV
(Fig.\,\ref{fig2}), while our calculations predict appearance of two
bands in this range. Their absence in the spectra is
easily understood because the $d-d$ transitions in
Fe$^{3+}$ compounds are spin-forbidden. Therefore, we
expect these bands to be too weak for detection in the samples as thin
as 50\,nm. An estimate gives optical density $D$ for these bands to be of the order of $10^{-2}$, even assuming the absorption coefficient
for them being 10 times higher than the one for the
$^6A_{1g}\rightarrow\,^4T_{1g}$ transition in
perovskite orthoferrite TmFeO$_3$.\cite{Usachev-PhysSolSt2005} On one hand, two broad bands centered at
2.23 and 2.94\,eV, respectively, can be assigned to the
$d-d$ transitions $^6A'_{1}\rightarrow\,^4A''_{1}$ and
$^6A'_{1}\rightarrow\,^4A'_{1},\,^4A'_{2},\,^4E''$, respectively (Fig.\,\ref{Fig-SpectralLevels}(c)). On the other hand, from the numerous
studies of 3$d$ metal oxides it is known that the strength of some
$d-d$ absorption bands can be considerably enhanced due
to admixture of the electric-dipole-allowed charge transfer (CT)
transitions from the valence band formed predominantly by the
occupied $2p$ oxygen orbitals to the conduction band formed by
unfilled 3$d$ iron orbitals. In perovskite orthoferrites the typical
photon energy for the lowest CT transition is $\sim$3\,eV.\cite{Kahn,Pisarev-PRB2009} In hexagonal orthoferrites one
can expect a red-shift of the fundamental absorption edge due to the
somewhat shorter Fe$^{3+}$-O$^{2-}$ distances.
Therefore, the absorption bands at 2.27 and 2.97\,eV
may have a combined CF and CT origin. The absorption bands
at 3.77 and 5.08\,eV should be purely CT transitions from $2p$
oxygen to the 3$d$ iron orbitals. We add that no absorption lines
associated with the rare earth ions were detected.

\begin{figure}[t]
\includegraphics[width=8.0cm,angle=0]{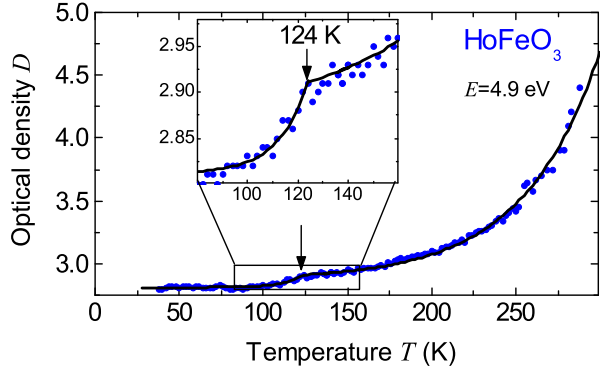}
\caption{(Color online) Temperature dependence of optical
density $D$ in HoFeO$_3$ at the photon energy $E=4.9$\,eV. Symbols show
experimental values and lines are exponential fit with different
parameters above and below $T=$124\,K.}\label{fig4}
\end{figure}

High-energy optical absorption of hexagonal $R$FeO$_3$ is
characterized by a sharp band centered at 5.08\,eV at $T=$293\,K. The absorption edge associated with this band is noticeably blue-shifted at $T$=20\,K (Fig.\,\ref{fig2}).
Fig.\,\ref{fig4} shows the temperature dependence of the absorption at 4.9\,eV in HoFeO$_3$. It is seen, that the absorption sharply decreases when
temperature is going down. A well defined anomaly is observed at 124\,K (see
inset). We relate the overall decrease of absorption in the temperature range 150-300\,K
to the freezing of phonons contributing to the broadening of the absorption band at 5.08\,eV. The anomaly at 124\,K in the temperature dependence of the absorption is in a very good agreement with recently reported temperature of the magnetic ordering for several hexagonal $R$FeO$_3$.\cite{Akbashev-APL2011,Jeong-JACS2012} Therefore, we assign the observed anomaly to the exchange splitting of ground and excited states
responsible for the given absorption band.\cite{Terasawa-JPhysC1980,Souchkov}

In conclusion, optical absorption was studied in thin films of multiferroic hexagonal rare earth
orthoferrites $R$FeO$_3$ ($R$=Ho, Er, Lu) grown
epitaxially on (111)-oriented YSZ substrates by MOCVD method. We demonstrate that optical absorption
spectra of hexagonal orthoferrites differ strongly from those of
perovskite orthoferrites and hexagonal manganites. Based on the analysis of the electronic structure in terms of the CF theory, we assign absorption
bands in hexagonal $R$FeO$_3$ at 2.27\,eV and 2.97\,eV to combined CF and CT transitions,
whereas the absorption bands at 3.77\,eV and 5.08\,eV to the
purely CT transitions from $2p$ oxygen to the 3$d$ iron orbitals. We add, that these results unveil the general features of the optical spectra and electronic structure of hexagonal $R$FeO$_3$ and, thus, may serve as a reference for more elaborated theoretical calculations, required for understanding various properties of these compounds.
Temperature dependence of absorption within the 5.08\,eV band shows
an anomaly near 124\,K, which is likely to originate from the magnetic ordering of Fe$^{3+}$ ions.

This work was supported by the RFBR (Projects No. 12-02-00130, 10-02-01008 and
10-03-00964), the FASI (Grant 02.740.11.0384) and the DFG. A. M. K. acknowledges support from The Committee for Science and Education of The Government of St. Petersburg in 2011.

\end{document}